# Fog based Secure Framework for Personal Health Records Systems

## Lewis Nkenyereye[1], S. M. Riazul Islam[2], Mahmud Hossain[3], M. Abdullah-Al-Wadud[4] and Atif Alamri[4]

[1]Department of Computer & Information Security, Sejong University, Seoul 05006, Korea
[2]Department of Computer Science and Engineering, Sejong University, Seoul 05006, Korea
[3]Department of Computer Science, University of Alabama at Birmingham, Alabama, 35294, USA
[4]Department of Software Engineering, King Saud University, Riyadh 11543, Saudi Arabia
[*]Corresponding Author: Lewis Nkenyereye. Email: nkenyele@sejong.ac.kr

**Abstract:** The rapid development of personal health records (PHR) systems enables an individual to collect, create, store and share his PHR to authorized entities. Health care systems within the smart city environment require a patient to share his PRH data with a multitude of institutions' repositories located in the cloud. The cloud computing paradigm cannot meet such a massive transformative healthcare systems due to drawbacks including network latency, scalability and bandwidth. Fog computing relieves the burden of conventional cloud computing by availing intermediate fog nodes between the end users and the remote servers. Aiming at a massive demand of PHR data within a ubiquitous smart city, we propose a secure and fog assisted framework for PHR systems to address security, access control and privacy concerns. Built under a fog-based architecture, the proposed framework makes use of efficient key exchange protocol coupled with ciphertext attribute based encryption (CP-ABE) to guarantee confidentiality and fine-grained access control within the system respectively. We also make use of digital signature combined with CP-ABE to ensure the system authentication and users privacy. We provide the analysis of the proposed framework in terms of security and performance.

## 1 Introduction

Personal health records (PHR) service consist of safely storing the patients records. These files contain the extensive summary of the patients medical history. With the advancement of the technology, the PHR management systems are predicted to allow a patient to create, manage, control and most importantly to share the PHR with a range of people and entities such as friends, medical doctors, nurses or any other authorized institution [1]. Thus, a patient can use his mobile device to perform all those operations. In 1996, the health insurance portability and accountability act (HIPAA) was released as an international law which requires a rigorous privacy and security for the patients and their PHR data. Lately, with the development of cloud computing, dedicated servers are outsourced by health agencies to store the PHR data [2]. Thus, causing several security, access control and privacy related concerns. In this regards, several access control protocol coupled with privacy preserving techniques were proposed both in industry and academia.

The most suggested technique to guarantee security, privacy and access control in healthcare environment is attribute based encryption (ABE). ABE can enable a patient to encrypt his/her PHR before storing them in outsourced cloud servers. The patient can specify through the ABE which person or entity is authorized to recover the PHR [3]. A typical example is a patient Bob who is recommended to take additional medical tests in a hospital which is different from the one he usually goes to. In order to proceed, Bob needs to share this existing PHR data to medical teams in that hospital. But Bob does not want to share his PHR data to the whole team, he wants to specify which data will be accessed by the nurses, the medical doctors or others. ABE will then allow the PHR users with corresponding secret keys to access the encrypted data since the decryptors in ABE need to have sufficient attributes to meet the designated access structure. ABE schemes are proposed in two main forms: KP-ABE and CP-ABE. In Key policy ABE (KP-ABE), the secret key of a given user are linked by an access structure set by an attribute authority (AA). However, in ciphertext policy ABE (CP-ABE), the ciphertext is directly linked with an access structure, which is set by the encryptor while the user secret keys are directly attached with a set of attribute. Thus, for an PHR scenario, the patient can first specify an access structure in ABE which a decryptor (PHR user) needs to satisfy before recovering the data.

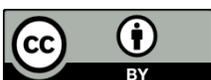

In this work, the concept of public domain (PUD) healthcare system was adopted. This notion assumes that the PHR users are very diversified. The PHR data might be needed by a range of authorized institutions beside the medical doctors and nurses. For instance, the patient can share his designated PHR data such as his height and weight to a gymnasium manager or any other sport based center. Thus, a need for a scalable and secure platform which can allow all the entities in PHR system to communicate is appealing. It is also worth mentioning that PHR data may include several media files such as pictures that would require sufficient computational capabilities.

Fog computing was introduced by Cisco as a scalable platform which would relieve the computational and communication burden from the remote servers located in the cloud and avails intermediate layer of fixed servers between the cloud servers and the IoT devices [4]. Several researchers have investigated on fog computing, specifically focusing on the quality of service (QoS) and quality of experience (QoS) . There are also a number of articles that investigated on security, privacy and access in fog based architectures. While the majority of the existing work focus on IoT, vehicular communications, smart grids and others; there are not pertinent work that address both access control, security, privacy preserving and fog-based scalability for the PHR systems. One of the relevant work in field was conducted by the researchers in [5]. They investigated pervasive health monitoring systems by achieving low latency and lightweight overhead for network congestion through fog computing. Their proposed scheme focused on patients that suffer from falls and strokes and provide a real time detection data analysis using fog computing. In this work, security, privacy, and access control concerns were not addressed specifically for PUD based healthcare systems.

Thus, looking at the security, privacy and access control requirements for PHR systems, the need of a scalable PHR system that can both offer security and quality of service within a PUD based healthcare system, the features of fog computing in terms of latency and lightweight network congestion, we are appealed to design a secure, privacy preserving protocol for public domain PHR over a fog -based architecture to address the afore mentioned limitations. The contribution of the proposed work is summarized as follows:

We first design an architectural model for secure and privacy preserving scheme for public domain PHR under a fog computing architecture. The proposed scheme allow the public domains entities in the PHR such as gymnasium to securely communicate with the PHR owner.

We design a secure protocol for public domain PHR over fog computing by combining the techniques of CP-ABE in order to offer security and access control coupled with digital signature technique.

We further proceed with the analysis of the proposed protocol in terms of security objectives, transmission overhead caused by the additional security primitives within the PHR and the simulation.

The remainder of this paper is structured as this. We first present the current state of art secure PRH systems, fog based computing architectures in section 2. In section 3, we outline the system model for the proposed scheme along with the basic security primitives for designing the proposed scheme. The proposed secure protocol is built in section 4.We conduct the security analysis and simulation performance of the proposed protocol in section 5 and draw the concluding remarks and future work in section 6.

**2 Related Work**

In the following section, we will discuss current literature that address security and access control in PHR systems and other existing work that proposed the adoption of fog computing in healthcare based systems.

*2.1 Attribute Based Encryption*

There is a considerable number of articles that proposed the combination of attribute based encryption along with other security primitive to enhance the security of systems. However, the majority of proposed protocol were suggested for the conventional cloud computing and Internet of Things (IoT) based applications such as mobile cloud computing [6-8] or vehicular cloud computing [9-11]. We focus in this section to review the state of art of the ABE based systems in PHR based applications.

In regards to PHR based systems, the researchers in [12] proposed a data sharing architecture that allow a patient to share his PHR data kept in the cloud by using an access control algorithm. In this work, they made use of ABE to ensure the privacy preservation of the patient identity and fine-grained access control for his PHR data. In [13], the authors proposed secure scheme for mobile healthcare cyber physical systems. The proposed scheme offered a number of countermeasure to protect resource limited mobile healthcare applications. The authors in [14] proposed a fined-grained access control scheme in mobile cloud based applications for healthcare domain. Their scheme combined a rigorous access control coupled with a mobile user authentication mechanism. There is also a number of articles that address privacy preservation for PHR data and access control for PHR users [15-17] but the majority of them do not build their secure protocols under a fog based architecture that address a diversified PUD healthcare systems.

*2.2 Fog Computing*

Introduced in 2012, fog based architectures or fog computing is a new paradigm that offer scalable solutions for applications and IoT devices by allowing them to run their applications in servers located in vicinity of the IoT device rather than remote servers placed in the cloud [18]. The fog computing is very crucial for the internet of things due to the number of devices which could be connected within a smart city. For instance, IoT deployments in mega smart cities is estimated to connect tens of billions devices which cannot always rely on the remote servers located in the cloud. Therefore, the fog paradigm is expected to decrease the network latency and congestion by offering efficient resource allocation solutions between the IoT devices and the remote servers. The fog based architectures have been extensively researched on especially in mobile based applications and IoT based scenario. The current literature that addressed fog computing focuses on three main issues: (1) the quality of service (QoS) which is deemed to be used in order to measure the performance of different entities such as the mobile operators within the fog architecture; (2) the quality of experience (QoE) was also widely researched on to depict the experience met by the end-users on top of the quality of service; (3) the security, privacy challenges and access control issues were researched on for a range of applications from mobile, cloud and IoT based environments. Nevertheless, the majority of the existing work focus on vehicular communications, smart grids and others [19-21], there are limited work that address both access control, security, privacy preserving and fog-based scalability for the PHR systems. One of the relevant work in the domain was conducted by the authors in [5]. They examined pervasive health monitoring systems by achieving low latency and lightweight overhead for network congestion through fog computing. Their proposed scheme mainly emphasized on patients that suffer from falls and strokes and provide a real time detection data analysis using fog computing. Though their work is a fog based model applicable healthcare environment, security, privacy, and access control concerns were not addressed specifically for PUD based healthcare system.

## 3 System Model and Preliminaries

In this section, we present first the system model of the proposed protocol which is made by the system architecture and the security goals which the proposed scheme has to satisfy. We then outline the basic security primitives used in the design of the proposed protocol including bilinear pairing and access structure.

### 3.1 Proposed System Model

The proposed system is a fog assisted architecture which is made by the following entities:

The attributes Authority: This is made by a number of servers located in the cloud with sufficient computing capabilities. These can be liken to companies or organization which work in the healthcare sector. Their main duties is to supervise how the PHR data are exchanged among the different entities. A typical example would be regional healthcare center (RHC) or any other departmental organization in healthcare domain.

The fog nodes: These are severs outsourced by the attributes authorities. Most of the time, the decryption can be done by the patient IoT device through the fog nodes. For instance, if the PHR contains several media files such as X-ray pictures, the patient can delegate the closest fog node to proceed with the recovery of the PHR file.

The PHR servers: These are databases in charge of storing encrypted PHR data belonging to PHR owners. These data are encrypted using the proposed protocol which is based on ABE.

The PHR owner: The data owner is typically any person or organization which has consulted a medical center for health related issues. The proposed protocol should enable a patient to securely create his data, manage his personal health data and most importantly control who is using his PHR data. In additional, the protocol should enable the patient to share his health data with other entities such as medical doctors from another hospital or any other authorized entity.

The PHR user: The PHR users also called PHR consumers are people or organization which need to access the patient's PHR data for treatment or any other health related issue. Assume a patient is located in city A and was recommended to continue his treatment in a hospital which is located in a city B. Our proposed protocol allow the PHR user (medical personnel in city B) to access the PHR data of the patient after his/her secured authorization.

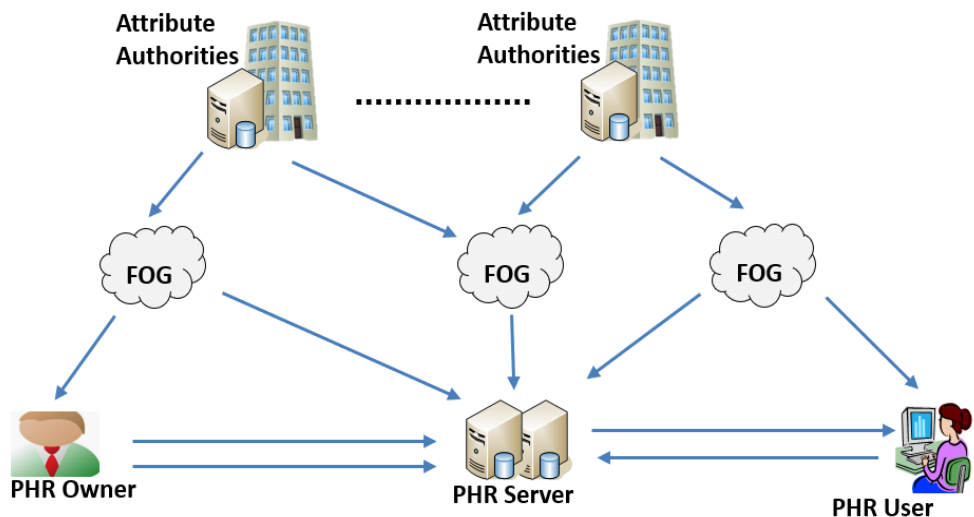

**Figure 1**: Proposed PHR Architecture

Fig. 1 depicts the proposed architecture which is made by the following entities: the attribute servers, the fog nodes, the PHR server, the PHR owner such as patient and the PHR consumer such as a medical doctor. In the following subsection, we provide the main security goals which the proposed framework needs to be satisfy.

*3.2 Security Goals*

The proposed protocol should be able to offer the following security goals both for the PHR owner, the PHR user, the fog nodes and PHR servers:

*Authentication*: The proposed scheme should be able to prevent any malicious user who wish to access any PHR data or learn any crucial information from the PHR data. Any entity, either a PHR user, PHR owner should be authenticated by the system before he gets involved in any communication.

*Confidentiality*: The proposed protocol should allow only legitimate PHR user (medical doctors or nurses,..) to access sensitive data of the PHR owner. The proposed protocol ensures the confidentiality of transmitted PHR data using specific security primitives.

*Fine-grained access control*: The proposed protocol should allow the PHR owner to specify which PHR user access which PHR data. For instance, the patient can specify that the sensitive medical test results can only be accessed by the medical doctors while other non-sensitive data such as weight or height can be accessed both by the medical doctors, their assistants or nurses.

*3.3 Preliminaries*

In this subsection we briefly described the two main security primitives which will be used in the proposed protocol. First we overview the basic concepts of bilinear pairing and then we describe the key notions of access structure which are used in attribute based encryption [22], [23].

*3.3.1 Bilinear Pairing*

Assume $G_1$ and $G_2$ to be two multiplicative cyclic group with a prime order $p$ and let $g$ be a generator of $G_1$. The bilinear map $e: G_1 \times G_1 \to G_2$ has to satisfy the following properties:

Bilinear: For all $w, s \in G_1$ and $a, b \in Z_p$, we have $e(w^a, s^b) = e(w, s)^{ab}$

Non-degenerate: The generator $g$ needs to satisfy $e(g, g) \neq 1$

Computable: For any $w, s \in G_1$, there is an efficient algorithm to generate $e(w, s)$.

*3.3.2 Access Structure*

Assume that we have a set of parties $\{D_1, D_2, \ldots, D_n\}$. We define a collection $\mathbb{A} \subseteq 2^{\{D_1, D_2, \ldots D_n\}}$ to be monotone if for $\forall B, C: B \in \mathbb{A}$ and $C \subseteq B$, that indicates that $C \in \mathbb{A}$.

A Monotone access structure is defined as monotone collection $\mathbb{A}$ if we have non empty subsets of $\{D_1, D_2, \ldots D_n\}$.

In a monotone collection $\mathbb{A}$, the sets within it are defined as authorized sets, and those that are not found in $\mathbb{A}$ are defined as unauthorized sets

In the proposed protocol , assume L to represent an access structure, with each non leaf node as a threshold gate, and each leaf node as an attribute. We node $num_x$ as the number of children of node $x$, and $k_x$ being the threshold value with $0 \leq k_x \leq num_x$. Note that the threshold value $k_x$ will return 1 if it is an OR gate or $num_x$ if it is an and gate. For every leaf node $x$, we set a threshold value of $k_x$ which is equals to 1. To make the access structure interpretation easy, we use the terms $parent(x)$ as the parent of the node $x$ within the tree, $att(x)$ as the attribute belonging to a leaf node $x$ and $index(x)$ being the function that outputs a unique value that is linked with a node $x$.

In the proposed scheme, a set of attributes act as a separate entity while a given access structure $\mathbb{A}$ specifies the policy within the attributes. Note that we mainly make use of monotonic access structures and simply labelled them as access structure in this work.

## 4 Proposed Protocol

The proposed secure protocol is designed within four main phase. In the first phase, an function is executed by the AA servers to generate the system public parameters and the master secret key. In the second phase, the AA servers also generate the secret keys based on a set of attributes S. The third main phase is executed by the PHR owner device which could be an IoT device or fixed computer. The PHR owner specifies the access tree structure that needs to be satisfied by the PHR user in order to recover the data. Finally, the last phase is performed by the PHR user in order to recover, note that in this phase, the PHR user can outsource the computing capabilities of the fog nodes to perform the decryption.

### 4.1 System Initialization

The first phase is executed by the AA server to generate the public parameters which are used both by the PHR user and the PHR owner. The details of the initialization function is described as follows:

$Initialization(U, T, 1^k) \rightarrow (MK, PK)$ : The different steps which are involved in the initialization phased are shown in Algorithm 1. The initialization function return the public parameters PK along with the master key MK. The public parameters PK can be accessed by all the entities in the system while the master key is securely key by attributes authority (AA).

**Algorithm 1** Initialization ($K$)

1: Select bilinear group $G_1$ and $G_2$ belonging to a prime order p and compute the generators $g_1$ and $g_2$

2: Select three random exponents $\alpha, \beta_1, \beta_2, \in Z_p$ such that $\beta_1 \neq \beta_2 \neq 0$

3: Select a hash function as a random oracle $H_1: \{0,1\}^* \rightarrow Z_p$

4: The public key is then computed as follows

$$PK = (G_1, G_2, g_1, g_2, p, H, h_1 = g_1^{\beta_1}, h_2 = g_1^{\beta_1}, e(g_1, g_1)^\alpha) \tag{1}$$

5: The master key is $MK = (\beta_1, \beta_2, g_1^\alpha)$

### 4.2 Private Key Generation

Prior to this stage, we assume that there is a secure communication between the AA server and the PHR server. The AA server will communicate a set of the attribute to the PHR server.

The AA defines the access structures $\mathbb{A}$ and forwarded securely to the PHR servers and fog nodes. For instance, a fog node can be given a set of attribute $fog_i=\{radiography, doctorlevel\ A, location\}$ and $fog_j=\{pediatry, nurselevel\ B, location\}$. Therefore, a fog node whose attribute satisfies the access structure can compute the shared key and recover the file.

In this phase, the function $KeyGen(PK, MK, S, T) \rightarrow SK$ is run to generate the private key. The detailed steps of the functions are shown in Algorithm 2. The PHR server then send the private key SK to the medical doctor, in this case the PHR user. Note that the proposed protocol adds a valid timing in which the private key can be used. For instance, if the patient has visited the hospital today, he can give a one-day or two days validity. We assume that $\mathbb{T}$ is minimum cover set of T, and $\mathbb{T}$ is set to contain several time intervals stated as $K = (\kappa_1, \kappa_2, ,,, \kappa_n)$. Therefore, practically, the private key can have a time validation such as $\mathbb{T}\{(2020,06,20), (2020,06,21), (2020,06,22)\}$ .

**Algorithm 2**: Key Generation *(MK, PK, S)*

1: Compute a key pair $(s_k, v_k)$ and choose a random $r, r \in Z_p$

2: Share $v_k$ to the rest of participants for verification

3: **for** Each $j \in S$ **do**
4:     Select $r_j \in Z_p$ and compute
5:     $D_j = g_1^r \cdot H(j)^{r_j}$ and $D_j' = g_1^{r_j}$
6: **end for**

7: The secret key $SK \in S$ is compute as $SK = (D = g_1^{\alpha + \frac{r}{\beta_1}}, E = g_1^{\frac{r}{\beta_2}}, \forall j \in S: D_j, D_j')$ (2)

### 4.3 PHR Encryption

The PHR encryption phase is shown in details within the Algorithm 3. In this phase, the patient which is the PHR owner upload the cypher text C which also contain symmetric key. The encryption can be performed by the PHR owner device or he can outsource the computing capabilities of fog nodes.

**Algorithm 3**: Encryption *(PK,T)*

1: Assume A to be an access structure characterized by T with root node R
2: Starting from the root R, then select a random $s \in Z$ and set $q_R(0) = s$
3: For each node $x$ in T, select a polynomial degree $q_x$ and set the degree to $d_x = k_x - 1$
4: **for** other nodes $x \in T$ **do**
5:     Set $q_x(0) = q_{parent(x)}(index(x))$
6:     Choose a random $d_x$ to define a polynomial $q\_x$
7: **end for**
8: Assume Y to be the set of leaf nodes in T with the verification key $v_k$, and let $K = e(g_1, g_2)^{\alpha s}$
9: The cyphertext is built as follows

$$CT = (T, C_1 = h_1^s, C_y = g_1^{q_y(0)})$$
$$C_y^1 = H(att(y))^{q_y(0)}, C_{v_k} = h_2^{q_{v_k}(0)}$$
$$C_{v_k}' = H(v_k)^{q_{v_k}(0)}: \forall y \in Y \quad (3)$$

10: Generate $\sigma = Sign_{s_k}(CT)$
11: The ciphertext is $C = (CT, \sigma)$

### 4.3 PHR Decryption

This phase allows the recovery of the PHR data by running the function $Decrypt(SK, PK, C) \to \perp$ or not, based on the attributes of the PHR user. Note that the *Lagrange theory* is used to recover the secret share. The decryption phase can executed by the PHR user devices (computer or mobile gadgets) or the PHR user can outsource the decryption phase to the fog nodes. In this phase, the PHR user or medical doctor first verifies the integrity of the PHR data along with the verification of the digital signature.

If everything holds, the PHR user proceeds by checking if the access time is still valid; if not the process will be aborted. Later on, the PHR user performs the decryption calculation by checking if the access policy set by the patient or PHR owner is satisfied. The PHR owner will compute the decryption key in order to recover the plaintext.

**Algorithm 4**: Decryption *(SK,PK, C)*

1: Verify the signature $\sigma$ using the key $v_k$

2: Compute $F_{v_k} = \frac{e(C_{v_k}, H(v_k) \cdot g_1^{\frac{r}{\beta_2}})}{e(C'_{v_k}, h_2)}$ (4)

3: **for** each node $x$ **do**
4:   **if** $x$ is a leaf node and $i \in S$ **then**
   $F_x = DecryptNode(CT, SK, x)$
5:     for all node $z$ that are children of $x$ do:
   $F_z = DecryptNode(CT, SK, z)$
6:   **end if**
7: **end for**
8: **if** $F_z \neq \perp$ **then**
   $F_x \prod_{z \in S_x} F_z^{\Delta_{i,S'_x}(0)}$, where $i = index(z)$
   $S'_x = index(z) : z \in S_x$
9: **end if**
10: **if** The node is a root R **then**
   $F_R = DecryptNode(CT, SK, R)$
11:   **if** $F_R == e(g_1, g_1)^{r \cdot q_R(0)}$ **then**
   $F_R = \prod_{x \in \{R, v_k\}} F_x^{\Delta_{index(x), \{R, v_k\}}}$
12:   **end if**
13: **end if**
14: Compute $\frac{e(C_1, D)}{A}$ to recover $K$

The correctness of equation 4 can be verified as follows:

$$F_{v_k} = \frac{e\left(C_{v_k}, H(v_k) \cdot g_1^{\frac{r}{\beta_2}}\right)}{e(C'_{v_k}, h_2)}$$

$$= \frac{e\left(C_{v_k}, g_1^{\frac{r}{\beta_2}}\right) \cdot e(C_{v_k}, H(v_k))}{e(C'_{v_k}, h_2)}$$

$$= \frac{e\left(h_2^{q_{v_k}(0)}, g_1^{\frac{r}{\beta_2}}\right) \cdot e(e(h_2^{q_{v_k}(0)}, H(v_k))}{e(H(v_k)^{q_{v_k}(0)}, h_2)}$$

$$= e(g_1^{\beta_2 \cdot q_{v_k}(0)}, g_1^{r/\beta_2})$$

$$= (g_1, g_1)^{r q_{v_k}(0)} \quad (5)$$

Then the function of decryption *Decryption (SK,PK, C)* is executed on the root R under the access tree T. The decryption process to recover the symmetric key is computed as follows:

$$K' = \frac{e(C_1, D)}{A} = \frac{e\left(h_1^s, g_1^{\alpha + \frac{r}{\beta_1}}\right)}{e(g_1, g_1)^{rs}}$$

$$= \frac{e(g_1,g_1)^{s(\alpha+r)}}{e(g_1,g_1)^{rs}}$$
$$= e(g_1,g_1)^{\alpha s} = K \tag{6}$$

## 5 Performance Analysis

In this section, we proceed with the evaluation of the proposed protocol in terms of security goals, computational cost caused by the message size and simulation.

### 5.1 Security Analysis

*Authentication*: Let us suppose that a malicious user probably another patient in the hospital wants to impersonate a legitimate PHR owner. In the scenario the malicious user needs to generate a valid private key within the Algorithm 2, $SK = (D = g_1^{\frac{\alpha+r}{\beta_1}}, E = g_1^{\frac{r}{\beta_2}}, \forall j \in S: D_j, D_j')$, which is not hard due to the unforgeability of CP-ABE scheme. Therefore, the scheme can provide authentication to all the entities.

*Confidentiality*: The PHR message is encrypted to generate the ciphertext $CT = (T, C_1 = h_1^s, C_y = g_1^{q_y(0)}$. In addition, a digital signature is performed on the encrypted data as $\sigma = Sign_{s_k}(CT)$. In the case, only a legitimate PHR owner can generate a valid signature, otherwise the decryption phase which is provided in equation 5 will not hold, therefore, the message will be discarded. Additionally, even if a malicious user eavesdrop the PHR data, only the hash value of the data will be obtained as shown in Algorithm 3.

*Fine-grained access control*: The access control is major security in the scenario. The patient which is the data owner might need to share his PHR to a medical team based on their rank, medical doctor or nurses. As shown in Algorithm 3, the ciphertext is constructed based on the access structure T as $CT = (T, C_1 = h_1^s, C_y = g_1^{q_y(0)})$. Therefore, unless a PHR user either a doctor or a nurse possesses valid secret shares to reconstruct the root node R, the decryption message will output $\perp$ as demonstrated in Algorithm 4 and equation 5.

### 5.2 Message Size

In this section, we proceed with the analysis of the message size of the proposed PHR scheme. As shown in the Algorithm 1 for the initialization process, the algorithm outputs the public key $PK = (G_1, G_2, g_1, g_2, p, H, h_1 = g_1^{\beta_1}, h_2 = g_1^{\beta_1}, e(g_1,g_1)^\alpha)$ along with the master key $MK = (\beta_1, \beta_2, g_1^\alpha)$. The outcome in total size is $|G_1| + |G_2| + |g_1| + |g_1| + |p| + |H| + |h_1| + |h_2| + |e| + |\beta_1| + |\beta_2| + |g^\alpha| = 11|G_1| + H$. In the second algorithm which is key generation, the secret key is made by $SK = (D = g_1^{\frac{\alpha+r}{\beta_1}}, E = g_1^{\frac{r}{\beta_2}}, \forall j \in S: D_j, D_j')$, which results in total message size of $|D| + |E| + ||S||(|D_j| + |D_j'|) = 2|G_1| + 2||s|||G_1| = (2 + 2||S||)|G\_1|$ with $|S|$ being the number of attributes. In the Algorithm 3 which is the encryption process, the output is the cipher text C along with the digital signature $\alpha$. Therefore the message size of $C = (CT, \sigma)$ is $|C_T| + |\alpha| = |C|$. In the last phase which Algorithm 4, the PHR user needs to keep the symmetric key K, therefore the message size is $|K|$. Note that the proposed scheme does not rely on certificate for authorization. Therefore, the computation overhead resulting from the use of certificate is generally higher, up to 8000 ms for the category of X.509 certificate while the size of the certificate can weight to tens of megabytes. In addition, the use of certificate to bind the identity of PHR owner to their keys cause an overhead within the transmission cost. The verification of the certificate validity and authenticity increase the overall time used in the decryption phase.

*5.2 Implementation*

In this section we proceed with the implementation of the proposed scheme which mainly focus on the impact of the number of attributes used in the access policy. We consider a range of 5 to 40 attributes for the implementation. We run our experiment using Python as the programming language under a Windows operating system version 10 Home, with processing capacity of 3.6 GHz, Intel core i7.

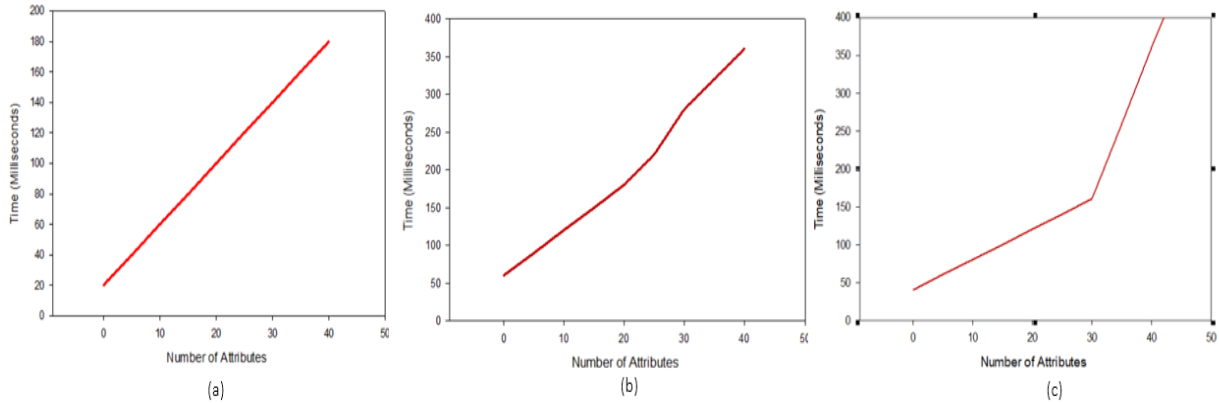

**Figure 2**: (a) Performance of the Key Generation Phase, (b) Performance of Encryption Algorithm and (c) Performance of the Decryption Algorithm

We also make use of the well-known and open source library for cryptographic operations Charm, which contains the PBC (standard pairing based cryptography) needed to execute mathematical operations for pairing based protocols. We first investigate the run time occurred on the key generation algorithm based on the described range of the attribute. As shown in Fig. 2(a), the run time is very linear, with the execution time increasing based on the number of attribute. We also investigate the performance of the encryption phase of the PHR by the patient. As depicted in Fig. 2(b), the polynomial operations caused by the child nodes in the access structure cause additional running time, but which is very limited in general. We later investigate the performance of the decryption phased which is illustrated in algorithm 4. As shown in Fig. 2(c), for a total of 25 attributes, the decryption time is less than 100 ms, which is acceptable for healthcare based applications looking at the security primitives within the proposed protocol.

**6 Conclusion**

In this paper, we proposed a secure and fog assisted framework for personal health records systems. The underlying fog architecture is used to allow public domain healthcare systems with a multitude of centers to use the computational capabilities of fog nodes for secure communications. The proposed secure protocol made use of ciphertext attribute based encryption to achieve confidentiality and rigorous access control for the PHR data. The CP-ABE was also coupled with digital signature to provide authentication of the all the entities within the system. We also provided the discussion of the proposed framework in terms of security along with the computational overhead. We further provided the implementation of the proposed protocol to investigate the impact of the security primitives on the system based on range of attribute, which yielded satisfactory results.

**Funding Statement:** The authors are grateful to the Deanship of Scientific Research at King Saud University for funding this work through Vice Deanship of Scientific Research Chairs: Chair of Pervasive and Mobile Computing.

**Conflicts of Interest:** The authors declare that they have no conflicts of interest to report regarding the present study.